\documentstyle[twoside]{article}

\catcode`\@=11
\long\def\@makefntext#1{
\protect\noindent \hbox to 3.2pt {\hskip-.9pt  
$^{{\eightrm\@thefnmark}}$\hfil}#1\hfill}		

\def\@makefnmark{\hbox to 0pt{$^{\@thefnmark}$\hss}}	
	
\def\ps@myheadings{\let\@mkboth\@gobbletwo
\def\@oddhead{\hbox{}
\rightmark\hfil\eightrm\thepage}   
\def\@oddfoot{}\def\@evenhead{\eightrm\thepage\hfil
\leftmark\hbox{}}\def\@evenfoot{}
\def\sectionmark##1{}\def\subsectionmark##1{}}

\oddsidemargin=\evensidemargin
\newcounter{sectionc}\newcounter{subsectionc}\newcounter{subsubsectionc}
\renewcommand{\section}[1] {\vspace{12pt}\addtocounter{sectionc}{1} 
\setcounter{subsectionc}{0}\setcounter{subsubsectionc}{0}\noindent 
	{\tenbf\thesectionc. #1}\par\vspace{5pt}}
\renewcommand{\subsection}[1] {\vspace{12pt}\addtocounter{subsectionc}{1} 
	\setcounter{subsubsectionc}{0}\noindent 
	{\bf\thesectionc.\thesubsectionc. {\kern1pt \bfit #1}}\par\vspace{5pt}}
\renewcommand{\subsubsection}[1] {\vspace{12pt}\addtocounter{subsubsectionc}{1}
	\noindent{\tenrm\thesectionc.\thesubsectionc.\thesubsubsectionc.
	{\kern1pt \tenit #1}}\par\vspace{5pt}}
\newcommand{\nonumsection}[1] {\vspace{12pt}\noindent{\tenbf #1}
	\par\vspace{5pt}}

\newcounter{appendixc}
\newcounter{subappendixc}[appendixc]
\newcounter{subsubappendixc}[subappendixc]
\renewcommand{\thesubappendixc}{\Alph{appendixc}.\arabic{subappendixc}}
\renewcommand{\thesubsubappendixc}
	{\Alph{appendixc}.\arabic{subappendixc}.\arabic{subsubappendixc}}

\renewcommand{\appendix}[1] {\vspace{12pt}
        \refstepcounter{appendixc}
        \setcounter{figure}{0}
        \setcounter{table}{0}
        \setcounter{lemma}{0}
        \setcounter{theorem}{0}
        \setcounter{corollary}{0}
        \setcounter{definition}{0}
        \setcounter{equation}{0}
        \renewcommand{\thefigure}{\Alph{appendixc}.\arabic{figure}}
        \renewcommand{\thetable}{\Alph{appendixc}.\arabic{table}}
        \renewcommand{\theappendixc}{\Alph{appendixc}}
        \renewcommand{\thelemma}{\Alph{appendixc}.\arabic{lemma}}
        \renewcommand{\thetheorem}{\Alph{appendixc}.\arabic{theorem}}
        \renewcommand{\thedefinition}{\Alph{appendixc}.\arabic{definition}}
        \renewcommand{\thecorollary}{\Alph{appendixc}.\arabic{corollary}}
        \renewcommand{\theequation}{\Alph{appendixc}.\arabic{equation}}
        \noindent{\tenbf Appendix \theappendixc #1}\par\vspace{5pt}}
\newcommand{\subappendix}[1] {\vspace{12pt}
        \refstepcounter{subappendixc}
        \noindent{\bf Appendix \thesubappendixc. {\kern1pt \bfit #1}}
	\par\vspace{5pt}}
\newcommand{\subsubappendix}[1] {\vspace{12pt}
        \refstepcounter{subsubappendixc}
        \noindent{\rm Appendix \thesubsubappendixc. {\kern1pt \tenit #1}}
	\par\vspace{5pt}}

\newcommand{\textlineskip}{\baselineskip=13pt}
\newcommand{\smalllineskip}{\baselineskip=10pt}

\def\eightcirc{
\begin{picture}(0,0)
\put(4.4,1.8){\circle{6.5}}
\end{picture}}
\def\eightcopyright{\eightcirc\kern2.7pt\hbox{\eightrm c}} 

\newcommand{\copyrightheading}[1]
	{\vspace*{-2.5cm}\smalllineskip{\flushleft
	{\footnotesize International Journal of Modern Physics A, #1}\\
	{\footnotesize $\eightcopyright$\, World Scientific Publishing
	 Company}\\
	 }}

\def\abstracts#1#2#3{{
	\centering{\begin{minipage}{4.5in}\baselineskip=10pt\footnotesize
	\parindent=0pt #1\par 
	\parindent=15pt #2\par
	\parindent=15pt #3
	\end{minipage}}\par}} 

\renewenvironment{thebibliography}[1]
	{\frenchspacing
	 \ninerm\baselineskip=11pt
	 \begin{list}{\arabic{enumi}.}
	{\usecounter{enumi}\setlength{\parsep}{0pt}
	 \setlength{\leftmargin 12.7pt}{\rightmargin 0pt} 
	 \setlength{\itemsep}{0pt} \settowidth
	{\labelwidth}{#1.}\sloppy}}{\end{list}}

\newcounter{itemlistc}
\newcounter{romanlistc}
\newcounter{alphlistc}
\newcounter{arabiclistc}

\newcommand{\fcaption}[1]{
        \refstepcounter{figure}
        \setbox\@tempboxa = \hbox{\footnotesize Fig.~\thefigure. #1}
        \ifdim \wd\@tempboxa > 5in
           {\begin{center}
        \parbox{5in}{\footnotesize\smalllineskip Fig.~\thefigure. #1}
            \end{center}}
        \else
             {\begin{center}
             {\footnotesize Fig.~\thefigure. #1}
              \end{center}}
        \fi}

\newcommand{\tcaption}[1]{
        \refstepcounter{table}
        \setbox\@tempboxa = \hbox{\footnotesize Table~\thetable. #1}
        \ifdim \wd\@tempboxa > 5in
           {\begin{center}
        \parbox{5in}{\footnotesize\smalllineskip Table~\thetable. #1}
            \end{center}}
        \else
             {\begin{center}
             {\footnotesize Table~\thetable. #1}
              \end{center}}
        \fi}

\def\@citex[#1]#2{\if@filesw\immediate\write\@auxout
	{\string\citation{#2}}\fi
\def\@citea{}\@cite{\@for\@citeb:=#2\do
	{\@citea\def\@citea{,}\@ifundefined
	{b@\@citeb}{{\bf ?}\@warning
	{Citation `\@citeb' on page \thepage \space undefined}}
	{\csname b@\@citeb\endcsname}}}{#1}}

\newif\if@cghi
\def\cite{\@cghitrue\@ifnextchar [{\@tempswatrue
	\@citex}{\@tempswafalse\@citex[]}}
\def\citelow{\@cghifalse\@ifnextchar [{\@tempswatrue
	\@citex}{\@tempswafalse\@citex[]}}
\def\@cite#1#2{{$\null^{#1}$\if@tempswa\typeout
	{IJCGA warning: optional citation argument 
	ignored: `#2'} \fi}}

\def\pmb#1{\setbox0=\hbox{#1}
	\kern-.025em\copy0\kern-\wd0
	\kern.05em\copy0\kern-\wd0
	\kern-.025em\raise.0433em\box0}

\def\fnt#1#2{\footnotetext{\kern-.3em
	{$^{\mbox{\scriptsize #1}}$}{#2}}}

\def\runninghead#1#2{\pagestyle{myheadings}
\markboth{{\protect\footnotesize\it{\quad #1}}\hfill}
{\hfill{\protect\footnotesize\it{#2\quad}}}}
\font\tenrm=cmr10
\font\tenit=cmti10 
\font\tenbf=cmbx10
\font\bfit=cmbxti10 at 10pt
\font\ninerm=cmr9

\font\eightrm=cmr8

\def\qed{\hbox{${\vcenter{\vbox{			
   \hrule height 0.4pt\hbox{\vrule width 0.4pt height 6pt
   \kern5pt\vrule width 0.4pt}\hrule height 0.4pt}}}$}}

\begin{document}

\def\beq{\begin{equation}}
\def\eeq{\end{equation}}
\def\beqa{\begin{eqnarray}}
\def\eeqa{\end{eqnarray}}
\def\be{\begin{equation}}
\def\ee{\end{equation}}
\def\bea{\begin{eqnarray}}
\def\eea{\end{eqnarray}}

\runninghead{Dimension eight effects...} {Dimension eight effects...}

\normalsize\textlineskip
\thispagestyle{empty}
\setcounter{page}{1}

\copyrightheading{}			

\vspace*{0.88truein}

\centerline{\bf NEW INSIGHTS CONCERNING DIMENSION EIGHT EFFECTS}
\vspace*{0.015truein} \centerline{\bf IN WEAK DECAYS}

\vspace*{0.37truein}
\centerline{\footnotesize JOHN F. DONOGHUE}
\vspace*{0.015truein}
\centerline{\footnotesize\it Physics Department, University
of Massachusetts, Amherst MA 01003}

\vspace*{0.21truein}
\abstracts{Most past work on weak nonleptonic decays has 
mixed dimensional regularization in the weak operator product 
expansion with some form of a cutoff regularization in the 
evaluation of the matrix elements. 
Even with the usual technique of matching the two schemes,
this combination misses
physics at short distance which can be described by 
dimension eight (and higher dimension) operators. I describe
some recent work with V. Cirigliano and E. Golowich which
clarifies these effects and provides a numerical 
estimate suggesting that they are important.}{}{}


\vspace*{1pt}\textlineskip	
\section{Introduction}	
\vspace*{-0.5pt}
\noindent

Weak nonleptonic decays are especially difficult to calculate. The
short distance nature of the W propagator implies that 
all values of the momentum from low energy up to the W mass
contribute to the process. Because we don't have a satisfactory method to
calculate all these scales at once, we attack the problem in parts,
using two ideas originally due to Ken Wilson - the separation of scales and 
the operator product expansion. We first imagine separating the problem into 
short distance physics and long distance physics, with the separation scale 
being called $\mu$. When we look at the short distance physics, we 
see the weak current product modified by the radiative corrections
of quarks and gluons. Since all aspects are at short distance we can 
represent the effect of these corrections as local operators. This 
brings in the operator product expansion, with the effective weak
Hamiltonian at this scale being given by a sum over a complete
set of operators with coefficients calculable in QCD perturbation
theory. However, we then need to add back in the long distance physics
by taking the matrix elements of the operators including physics
of energies up to the scale $\mu$. Such matrix elements cannot use 
perturbation theory but are best accomplished with low energy hadronic 
methods. This separation of the physics into long and short distance
regions using the OPE allows us to use the best methods available
for each region.

The above description is what we as a field have always thought that
we were doing both in theory and in practice. However, in fact most
calculations were not doing this. The departure from the above
description comes in the use of dimensional regularization in the
calculation of the OPE coefficients. Dimensional regularization does
not provide a complete separation of scales. The parameter $\mu$ that 
appears in dimensional regularization does not function as a boundary
between short and long distance. In practice, only the most singular
pieces of short distance physics is captured when regularizing 
dimensionally.
This is sufficient to regularize the singular local operators of the 
OPE and to sum the large logarithms of the W mass. But it does not
provide the complete Wilsonian integrating-out of all short distance
physics. Some less singular physics remains to be included in the
physical amplitudes.

Of course, dimensional regularization is a perfectly fine scheme to 
use in the calculation, if 
employed consistently. But in practice most calculations
mix regularization schemes, using some form of a cutoff to calculate 
matrix elements. Thus the matrix elements are done with a rigorous
separation of scales, while the coefficients are not. This is not
consistent. What is missing
from the final answer is the less singular short distance physics 
from the short distance parts of the calculation. 

This conflict can be phrased in terms of the operators that appear
on the OPE. Dimensional regularization involves operators of
dimension six in the OPE. The less singular
short distance physics can be described by operators of dimension 
eight and higher. The question is where the physics of 
dimension eight appears. Present practice 
must be modified in one of the following ways:
\begin{itemize} 
\item If one calculates the matrix elements 
including physics only up to the scale $\mu$,
one must include dimension-eight operators in 
the OPE with coefficient of order
$1/\mu^2$.
\item If instead one wants to use dimensional regularization 
throughout, one must 
include physics of all scales in the matrix elements. The high energy 
portion of the matrix element above the 
scale $\mu$ is the effect of dimension-eight
(and higher) operators.
\end{itemize}
\noindent
Our paper$^1$ provides a concrete demonstration of both of these
options and a calculation of the dimension eight operators
relevant for the Standard Model at one loop. Here I can merely sketch 
the flow of this demonstration.

We calculate a particular type of weak matrix element that
through some manipulation can be written as the integral over
vector and axial vacuum polarization functions.
\beq
{\cal M}  =   {3 G_F M_W^2 \over 32 \sqrt{2}\pi^2 
F_\pi^2} \int_0^{\infty} dQ^2 \ {Q^4 \over Q^2 + M_W^2}   
        \left[ \Pi_{V,3} (Q^2) - \Pi_{A,3} (Q^2) \right] \ \ .
\label{full}
\eeq
Here the hadronic information is contained in the vacuum polarization
functions. We can pick out all the ingredients to the amplitude
by studying this function.

  First we will consider the case where we separate the
physics above and below some value of $Q^2=\mu^2$.
The high energy parts of the vacuum polarization can be written
(in the chiral limit)
in terms of operators of dimension-six, dimension-eight and higher.
The specific form is not too interesting here, but the reader should
just follow the superscript to the operator, which indicates the dimension.
Thus
\beq
(\Pi_{V,3} - \Pi_{A,3})(Q^2) \sim 
{2 \pi \langle \alpha_s {\cal O}^{(6)}_8 \rangle_{\mu} \over Q^6} + 
 {{\cal E}^{(8)}_\mu \over Q^8} + \dots \ \ .
\label{c2}
\eeq
To complete the calculation in the scheme where we fully separate the scale
we define a hadronic matrix element by including
all physics up to an energy cutoff at the scale $\mu$. We end up 
finding
\beq
{\cal M}  \simeq  {G_F \over 2 \sqrt{2}F_\pi^2}
\left[\langle {\cal O}^{(6)}_1 \rangle_\mu^{\rm (c.o.)} 
+ \ { 3 \over 8 \pi} \ln \left( {M_W^2 \over \mu^2}
\right) \langle \alpha_s {\cal O}^{(6)}_8 \rangle_{\mu} \ 
+{3\over 16 \pi^2}{{\cal E}^{(8)}_\mu \over \mu^2} + \ldots \right]  .
\label{sd11}
\eeq
where the first two operators are of dimension six.
This is the correct OPE for this matrix element in the case where
we fully separate the short distance from long distance physics. In this
case the dimension eight operator appears in the OPE and has
come for the sub-leading effect at short distance.

On the other hand we can go back and define the local operator 
not by a cutoff but by dimensional 
regularization. This results in a definition in which all 
momentum scales are present 
\beq
{\langle {\cal O}^{(6)}_1 \rangle_{\mu_{\rm d.r.}}^{\rm (d.r.)} } 
=  { (d - 1) \mu_{\rm d.r.}^{4 -d} 
\over (4 \pi)^{d/2} \Gamma(d/2)} \int_0^\infty dQ^2 
\ ~Q^d  \left( \Pi_{V,3} - \Pi_{A,3} \right)(Q^2)              
\label{r2}
\eeq
Even without evaluating the integral we can
see that to know its value we must include physics from
above the scale $\mu_{\rm d.r.}$, since there is no
separation of scales.
After removing the divergences in the ${\overline {\rm MS}} $ scheme,
we can then obtain the renormalized operator
\beq
\langle {\cal O}^{(6)}_1 
\rangle_{\mu_{\rm d.r.}}^{\rm {\overline {\rm MS}}}
= \langle {\cal O}^{(6)}_1 \rangle_\mu^{\rm (c.o.)}
+ {3 \alpha_s \over 8 \pi} \left[ \ln\left({\mu_{\rm d.r.}^2 \over 
\mu^2}\right) - {1\over 6}\right] \langle {\cal O}^{(6)}_8 \rangle_\mu 
+ {3\over 16 \pi^2}{{\cal E}^{(8)}_\mu \over \mu^2} \ \ .
\label{va37}
\eeq
The mixing of dimension six operators is expected in a typical matching
situation. However, we also see that dimension eight operators need 
to be included in the matching between a cutoff scheme and dimensional
regularization. This is what is normally missed. It comes from the
part of the integral 
that is above the cutoff, i.e. from short distance physics. 
The particular size of 
the dimension eight component in the operator is exactly the same as 
was found in the OPE above. Therefore the OPE in a dimensional scheme
can be done with only dimension-six operators in the OPE. The dimension
eight effect is in the matrix element instead.

Within this calculation, we can also calculate reliably the magnitude of 
the dimension eight effect. This is because the vacuum polarization functions
satisfy dispersion relations with the input being 
given by data on $e^+e^-$ reactions
and $\tau$ decays. What we find is that the dimension
eight correction is 100\% 
of the leading dimension six operator when $\mu \sim 1.5$~GeV, and scales 
as $1/\mu^2$.

Most calculations of $\epsilon' / \epsilon$ reported in the literature
were done with scales at 1 GeV or below. These dimension eight
effects (and even higher dimension effects) 
certainly would be present. These constitute an extra
uncertainty in the predictions.

\nonumsection{References}

\end{document}